\documentclass[12pt]{article}
\usepackage[english]{babel}
\usepackage{graphicx}

\large \baselineskip = 20 true pt

\begin{document}
\begin{center}
{\Large \bf The Equivalent Conversions of the Role-Based Access Control Model} \vspace{0.5cm}
\end{center}

\begin{center}
N.F. Bogachenko\\
Dostoevsky Omsk State University, Omsk, Russia

 \vspace{0.5cm}
\end{center}

\begin{center}
{\bf Abstract}
\end{center}

The problems which are important for the effective functioning of an access control policy in a large information system (LIS) are selected. The general concept of a local optimization of a role-based access control (RBAC) model is formulated. The optimization criteria are proposed. The algorithms of a local optimization of the RBAC model are defined and justified. The developed algorithms are used in the methods of the solution of the following problems: the assessment of risks of the leakage of permissions in the RBAC policy, the access control in the distributed hierarchical systems, the combining of role-based and mandatory access control models. In the first problem the question of the permissions distribution in the role hierarchy is researched. The analytic hierarchy process (AHP) is applied to creation of the estimates. The method is based on the hierarchical structure of a role set. The offered technique can order the permissions according to the value of the risks of their leakage. In the second problem the algorithm of the distribution of the cryptographic keys in the system with a hierarchical arrangement of the objects is offered. The cryptography protocols for the practical use of this algorithm are defined. The conditions of the implementation of the discretionary and mandatory principles of the access control on the basis of the developed algorithm are formulated.


\section{Introduction}

The research objective is the development of methods and algorithms of the solution of the problems which arise in the case of the implementation of an access control policy in a LIS.  The specifics of the data access arrangement in a LIS are defined. Here are the most significant factors: 

1. The demand of the RBAC mechanisms as the object-oriented decision which is capable to reduce complexity of the LIS administration~\cite{b1, b2}. 

2. The need for the combination of the several access control models in the case of the continuous functioning of subsystems where these principles of the access control 
are implemented~\cite{b3, b4, b5}. 

Now interest in RBAC is shifted towards the role mining problem~\cite{b6, b7, b8, b9} and different "supermodels"~\cite{b10, b11}. 
One more direction of scientific activities is development of methods of RBAC configuration for cloud computing~\cite{b12, b13}.

Based on these factors a set of the problems which are important for the effective work of a LIS safety subsystem is selected.

1.	The creation and the supporting of the RBAC.

(1.1).	The deleting the duplicate roles. 

(1.2).	The taxonomic distribution of permissions.

(1.3).	The optimal file representation of a role hierarchy.

(1.4).	The assessment of risks of the leakage of permissions.

(2).	The access control on the basis of the computable cryptographic keys.

(3).	The combining of the different security models in a computer system.

The efficiency of the solution of these problems significantly depends on the structure of the role hierarchy which is used in the LIS access control policy. The additional properties of the role hierarchy which are required for the solution of the formulated problems are found: tree-like hierarchy; leaf hierarchy; $RP$-reduced hierarchy; transitive-reduced 
hierarchy (tab.~\ref{tab1})~\cite{b14, b15}.
\begin{table}[ht]
\centering
\caption{Characteristics of the role hierarchy which are necessary for the solution of some access control problems}
\label{tab1}
\begin{tabular}{|c|c|c|c|c|}
\hline
 \textbf{Role  hierarchy}     & tree-like & leaf & $RP$-reduced & transitive-reduced\\ 
\hline
\hline
\textbf{Number} &\multicolumn{2}{|c|}{(1.4)} & (1.1) &  \\
\cline{2-4}
\textbf{of the} & (2) & \multicolumn{2}{|c|}{(1.2)}  & (1.3)  \\
\cline{2-2}
\textbf{problem} & (3)& \multicolumn{2}{|c|}{ }  &  \\
\hline
\end{tabular}
\end{table}

There is a need for development of the conversion methods of the role hierarchy according to the specified characteristics. At the same time changes of an access control subsystem must be transparent for the user: a user is obliged to receive the same permissions before and after the conversions.

\section{Local optimization of the RBAC}

The graph-based representation of the RBAC model was formalized for more strict determination of the possible conversions of the role hierarchy. It is known that RBAC model is defined by the set of following elements: $U$, $P$, $R$ -- sets of users, permissions and roles; 
$RP: R \rightarrow 2^P$, $UR: U \rightarrow 2^R$, $RR: R \rightarrow 2^R$, $UP: U \rightarrow 2^P$ -- mappings which define the distribution of permissions between roles, the authorization of users for roles, the authorization of roles at each other, the users permissions. 

The directed marked graph which has no directed cycles (the \textit{role graph}) $G = (R, E, RP)$ is the graphical representation of sets $P$, $R$ and mappings $RP$, $RR$. The arc $(r_i, r_j)$ exists in the role graph if and only if $r_j \in RR(r_i)$.

RBAC models are named the \textit{equivalent} models if the sets of permissions coincide, the sets of users coincide too, and the user's permissions mappings are isomorphic in these models. 

Required conversions of the RBAC model must consist in the following:

1)	to lead the role hierarchy to the required look; 

2)	to lead to creation of the equivalent RBAC model; 

3)	to make the minimum changes to the main sets and mappings of the RBAC model. 

These conversions are named the \textit{local optimization} of the RBAC model.  By definition the local optimization comes down to conversions of a role graph.

For further formalization the $RP$-conversions of a role graph are defined. These conversions must lead to creation of the equivalent RBAC model. The conversion of a role graph $G$ to a role graph $G^*$ is $RP$-\textit{admissible} if the following conditions are satisfied:  

1) 	$RP(G) \subseteq RP(G^*)$; 

2)	$\forall$ directed path $\rho(r_i, r_j) \in G$ $\exists$ "conjugate"\ directed path 
$\rho(r_u, r_v) \in G^*$: $RP(r_i) = RP(r_u) \wedge RP(r_j) = RP(r_v)$. 

The conversion of a role graph $G$ to a role graph $G^*$ is $RP$-\textit{equivalent} if the following conditions are satisfied:

1)	 $RP(G) = RP(G^*)$; 

2)	the requirement of the existence of the "conjugate"\ directed path must be fulfilled both for an initial role graph and for a resultant role graph.

It is proved that the conversion $F$ of a role graph $G$ to a role graph $G^*$ is the local optimization of the RBAC model if the following conditions are satisfied: 

1)	$G^*$ meets the selected criterion of optimality; 

2)	$F$  is  $RP$-admissible (or $RP$-equivalent) conversion; 

3)	the number of the nodes and/or of the arcs of a role graph did not increase or this increase is minimum.

Four criteria of the local optimization of a role graph are selected: tree-like role graph, leaf role graph, $RP$-reduced role graph and transitive-reduced role graph. 

Some propositions and theorems which define and justify the algorithms of the local optimization of the RBAC model are proved~\cite{b14, b15}. These algorithms lead to creation of the equivalent RBAC model. Four main and four derivative algorithms are received (tab.~\ref{tab2}). 
\begin{table}[ht]
\centering
\caption{Algorithms of the local optimization of the RBAC model}
\label{tab2}
\begin{tabular}{|c|c|c|}
\hline
 \textbf{Algorism} & \textbf{Conversion} & \textbf{Features of the optimal role graph}\\ 
\hline
\hline
\multicolumn{3}{|c|}{Main algorithms}   \\
\hline
(I) & $RP$-admissible & single, leaf\\
\hline
(II) & $RP$-equivalent & $RP$-reduced\\
\hline
(III) & $RP$-equivalent & tree-like\\
\hline
(IV) & $RP$-equivalent & transitive-reduced\\
\hline
\multicolumn{3}{|c|}{Derivative algorithms}   \\
\hline
(Ia) & $RP$-admissible & leaf\\
\hline
(I+II) & $RP$-admissible & single, taxonomic, leaf, $RP$-reduced\\
\hline
(III+I) & $RP$-admissible & leaf,  tree-like\\
\hline
(III+Ia) & $RP$-admissible & single, leaf,  tree-like\\
\hline
\end{tabular}
\end{table}

The complexity of the constructed algorithms is a polynomial in the number of roles and permissions.
The algorithms of the local optimization of the RBAC model were used for creation of the methods of the solution of some access control problems (tab.~\ref{tab1}). The results received for problems (1.4) (2) and (3) follow.

\section{Applications of the algorithms of the local optimization}

The assessment problem of risks of the leakage of permissions in the RBAC policy is formalized: probability of leakage of each permissions $p_i \in P$ must be evaluated. The new algorithm of the solution of this problem uses AHP and the role graph of the RBAC policy. The tree-like leaf role hierarchy is necessary for implementation of the offered approach. This type of hierarchy is provided by algorithm (III+I). The advantages of the developed technique are defined and justified: 

1.	The "model"\ error of AHP which arises in a consequence of inconsistency of opinions of experts is removed. 

2.	The automation of the process of the ordering of the permissions according to the value of the risks of their leakage is possible.

The problem of the access control in the hierarchical system is defined: the set of the system's objects is partially ordered; access control model for this system must be defined~\cite{b16}. The execution of the following conditions is supposed:

1.	The object hierarchy is defined by the digraph $G$; the order $G$ is $n$.

2.	Each object $O_i$ ($i = 1, \ldots, n$) is encrypted by a secret key $k_i$  (an \textit{access} key); a symmetric encryption is used.

3.	An access to the object $O_i$ is possible if the access key $k_i$ is known and the condition of a sequential access is satisfied. The sequential access consists in the following: if a user wants to get an access to the object $O_i$, he must have an access to all objects which form the directed path from the hierarchy root to the object $O_i$.

In the local systems this problem can be solved by a security subsystem. In the distributed system a uniform security subsystem is absent; the access control is reached by means of algorithms of a distribution of cryptographic keys and an encryption. The new method of the distribution of the access keys is offered. This method is based on the principle of the computability of keys~\cite{b16}. For the implementation of this approach the object hierarchy must be tree-like. Algorithm (III) provides this requirement.

The new method of the combining of RBAC and mandatory access control (MAC) is offered. This approach is based on the search of the "ideal"\ solution: the access rules must meet requirements of the both models and must not contradict each other. The Cartesian product of the MAC lattice and the "role"\ lattice which is generated by the role hierarchy is considered. Algorithm (III) is involved for the receiving a "role"\ lattice.

Further problems (1.4) and (2) will be represented in more detail.

\section{The assessment of risks of the leakage of permissions}

The heuristic assumptions are formulated: 

1. The more the number of permissions of the role is, the more probability of the attack on this role is. 

2. The more the prevalence of the permission in the role hierarchy is, the more the probability of the leakage of this permission is. 

3. The more the distance of the role from the hierarchy root is, the less the probability of the attack on this role is.

According to algorithm (III+I) of a local optimization of the RBAC model we can assume that the role graph is the tree-like leaf digraph $T$. 

\begin{figure}[ht]
\centering
\includegraphics[width=0.6\textwidth]{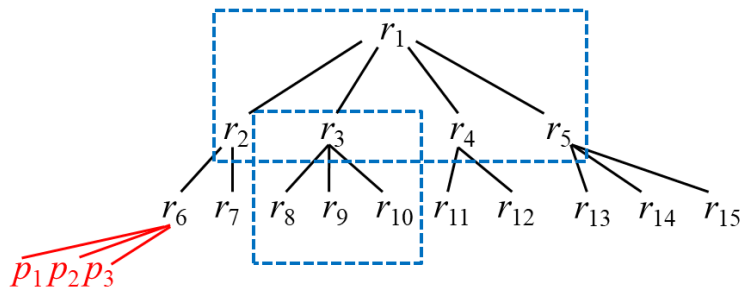}
\caption{The fragment of the extended tree $T_p$}
\label{fig1}
\end{figure}

At the preparatory stage the tree $T$ extends to the tree $T_p$: to each leaf $r_l$ of the tree $T$ the new nodes are added; the new node contains one permission from the set of permissions $RP(r_l)$ of the leaf $r_l$ (fig.~\ref{fig1}). The tree of the solution of the AHP is the tree $T_p$.

At the first stage the relative coefficients of all nodes (except a root) of the tree $T_p$ are calculated. The calculation of the coefficients takes place in the direction from the root to the leaf. The nodes $r_{s1}, \ldots, r_{sk}$ which are subordinated to one node $r_s$ from the previous level are considered. In terms of AHP the selected nodes $r_{s1}, \ldots, r_{sk}$ are the alternatives for the criterion $r_s$. For each subset $\{r_{s1}, \ldots, r_{sk}\}$ the paired comparisons matrix $\textbf{M}_s$ is built. This matrix is used for the calculation of the relative coefficients of the nodes $r_{s1}, \ldots, r_{sk}$. The matrix element $[\textbf{M}_s]_{ij}$ which corresponds to pair $(r_{si}, r_{sj}$) is equated to the relation of the number of permissions of the role $r_i$  to the number of permissions of the role $r_j$: 
$$[\textbf{M}_s]_{ij} = \frac{\vert RP(r_{si}) \vert}{\vert RP(r_{sj}) \vert}.$$

It is proved that these paired comparisons matrixes are ideally coordinated.  As a result the relative coefficient $w_{si}$ of the node $r_{si}$ is calculated according to the formula:
$$w_{si} = \frac{\vert RP(r_{si}) \vert}{\sum_{j=1}^k \vert RP(r_{sj}) \vert} .$$

At the second stage the combined coefficients (probabilities or risks of the leakage of permissions) are calculated. The probability of the leakage of the permission $p$ is equal to the sum over all directed paths from the root to the leaves which contain this permission. Each item of the sum is the product of the relative coefficients of the nodes which form one of the directed paths.
$$P(p) = \sum_{\rho(root,\ leaf):\ RP(leaf)=\{p\}} \left( \prod_{j:\ r_j\in\rho(root,\ leaf)} w_j \right) .$$

The complexity of the suggested algorithm is a polynomial in the number of roles $n$ and permissions $m$: T = O($(n \times m)^2$).

\section{The access control on the basis of the computable cryptographic keys}

One secret key $k_0$ is determined and stored. All access keys are calculated one of another according to an object hierarchy: if an arc $(O_i, O_j)$ is in the hierarchy then the access key $k_j$ of the object $O_j$ is a value of the function $h$, which is infeasible to invert ($h$ is one-way function). The function $h$ depends on the access key $k_i$ of the object $O_i$ and the properties of the object $O_j$: $k_j = h(k_i, O_j)$. The algorithm of the formation of computable access keys for a tree-like object hierarchy is defined. This algorithm uses cryptographic hash function as function $h$.

1.  The unique identifier $id_i$ is assigned to each object $O_i$. The general initialized key $k_0$ is defined. 

2. The access key $k_1$ of the object $O_1$ which is a root of the object hierarchy is calculated by the formula: $k_1 = h(k_0 \circ id_1)$, where $(k_0 \circ id_1)$ is the concatenation of the key and the identifier. 

3. For each arc $(O_i, O_j)$ of the object hierarchy the access key $k_j$ of the object $O_j$ is calculated by the formula:  
$$k_j = h(k_i \circ id_j).$$

The suggested algorithm is named "hash-based access keys distribution"\ (HBAKD). It is proved that existence of the tree-like object hierarchy is necessary and sufficient condition of uniqueness of the computation of access keys according to the HBAKD algorithm. The generalized algorithm of the formation of computable access keys for arbitrary object hierarchy which is described by the digraph $G$ is defined:

1.  The directed tree $T$ which is $ID$-\textit{equivalent} to the digraph $G$ is constructed. For this purpose the modification of algorithm (III) of a local optimization of the RBAC model is used. 

2. According to the HBAKD algorithm the access keys for all nodes of the tree $T$ are computed.

3. Secret sharing scheme can be applied to each $ID$ class of the tree $T$ if this class consists of several elements.

\begin{figure}[ht]
\centering
\includegraphics[width=0.7\textwidth]{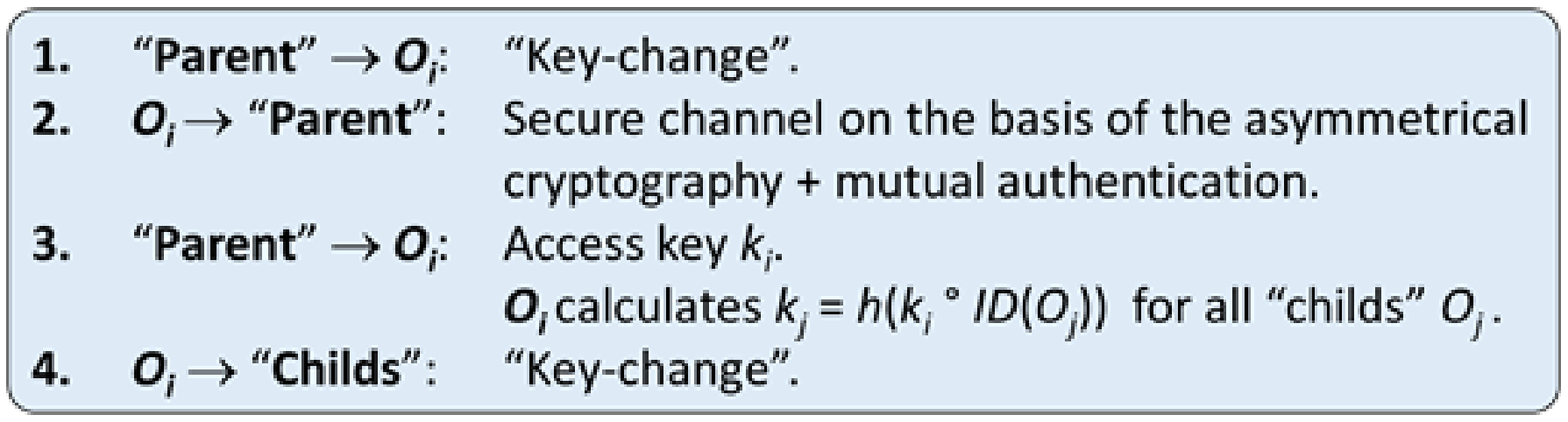}
\caption{The scheme of "Key-change"\ protocol}
\label{fig2}
\end{figure}

The cryptographic protocols for the practical use of the HBAKD algorithm are offered. The main one is "Key-change"\ protocol (fig.~\ref{fig2}). The possible attacks on these protocols are analyzed. The additional safety measures which are put in protocols are proved (tab.~\ref{tab3}).

\begin{table}[ht]
\centering
\caption{The additional safety measures of the "key-change"\ protocol}
\label{tab3}
\begin{tabular}{|c|c|}
\hline
 \textbf{Possible attacks} & \textbf{Safety measures}\\ 
\hline
Interception of the message "Key-change"; substitution of the "child" & Step 2 \\
\cline{1-1}
Sending the false message "Key-change"; substitution of the "parent" &  of the protocol\\
\hline
\end{tabular}
\end{table}

The conditions of the implementation of the discretionary and mandatory principles of the access control on the basis of the HBAKD algorithm are formulated. It is supposed that the set of the objects is partially ordered and the access to any object is possible only sequentially from the "parent"\ object to the "child"\ object. It is proved that the implementation of the mandatory and discretionary models of the access control is possible for a tree-like object hierarchy in the case of the execution of the conditions which are listed in table~\ref{tab4} ($ID_S$ is the set of the identifiers of the objects which are accessible for the subject $S$).

\begin{table}[ht]
\centering
\caption{Conditions of the implementation of the discretionary and mandatory principles of the access control}
\label{tab4}
\begin{tabular}{|c|c|c|}
\hline
 \textbf{Access control} & \textbf{Identifiers} & \textbf{It is known to the $S$}\\ 
\hline
Mandatory & Open & $k_i$\\
\hline
Discretionary & Secret & $k_0$ + $ID_S$\\
\hline
Mandatory and  discretionary & Secret & $k_i$ + $ID_S$\\
\hline
\end{tabular}
\end{table}

\section{Conclusions}

Some problems which are important for the effective work of an access control policy in a LIS have been discussed. The solution of these problems significantly depended on the structure of the hierarchy of the system's entities. A special attention has been paid to the RBAC model. The general concept of the local optimization of the RBAC model has been formulated. This process consists in the equivalent conversions of a role graph. The equivalent conversions make the minimum changes to the main sets and mappings of the RBAC model and increase the efficiency of the functioning of the system due to the coercion of the role hierarchy to the required look. The tools which can execute the specified conversions of the RBAC model have been developed. On the basis of optimization algorithms the methods of the creation and of the supporting of the access control policy in a LIS have been obtained. The correctness of the offered approaches, methods and algorithms has been confirmed by the rigorous mathematical proofs and by the results of the computing experiments.


\end{document}